\documentclass[prl,twocolumn,aps]{revtex4}
\usepackage{graphicx}
\usepackage{epsfig}

\begin{document}


\title{Quantum Communication Through an Unmodulated Spin Chain}

\author{Sougato Bose}

\affiliation{Institute for Quantum Information, MC 107-81,
California Institute of Technology, Pasadena, CA 91125-8100, USA}
\affiliation{Department of Physics and Astronomy, University
College London, Gower St., London WC1E 6BT, UK}



\begin{abstract}
We propose a scheme for using an unmodulated and unmeasured
spin-chain as a channel for short distance quantum communications.
The state to be transmitted is placed on one spin of the chain and
received later on a distant spin with some fidelity. We first
obtain simple expressions for the fidelity of quantum state
transfer and the amount of entanglement sharable between any two
sites of an arbitrary Heisenberg ferromagnet using our scheme. We
then apply this to the realizable case of an open ended chain with
nearest neighbor interactions. The fidelity of quantum state
transfer is obtained as an inverse discrete cosine transform and
as a Bessel function series. We find that in a reasonable time, a
qubit can be directly transmitted with better than classical
fidelity across the full length of chains of up to 80 spins.
Moreover, the spin-chain channel allows distillable entanglement
to be shared over arbitrarily large distances.
\end{abstract}


\maketitle

\begin{figure}
\begin{center}
\includegraphics[width=3in, clip]{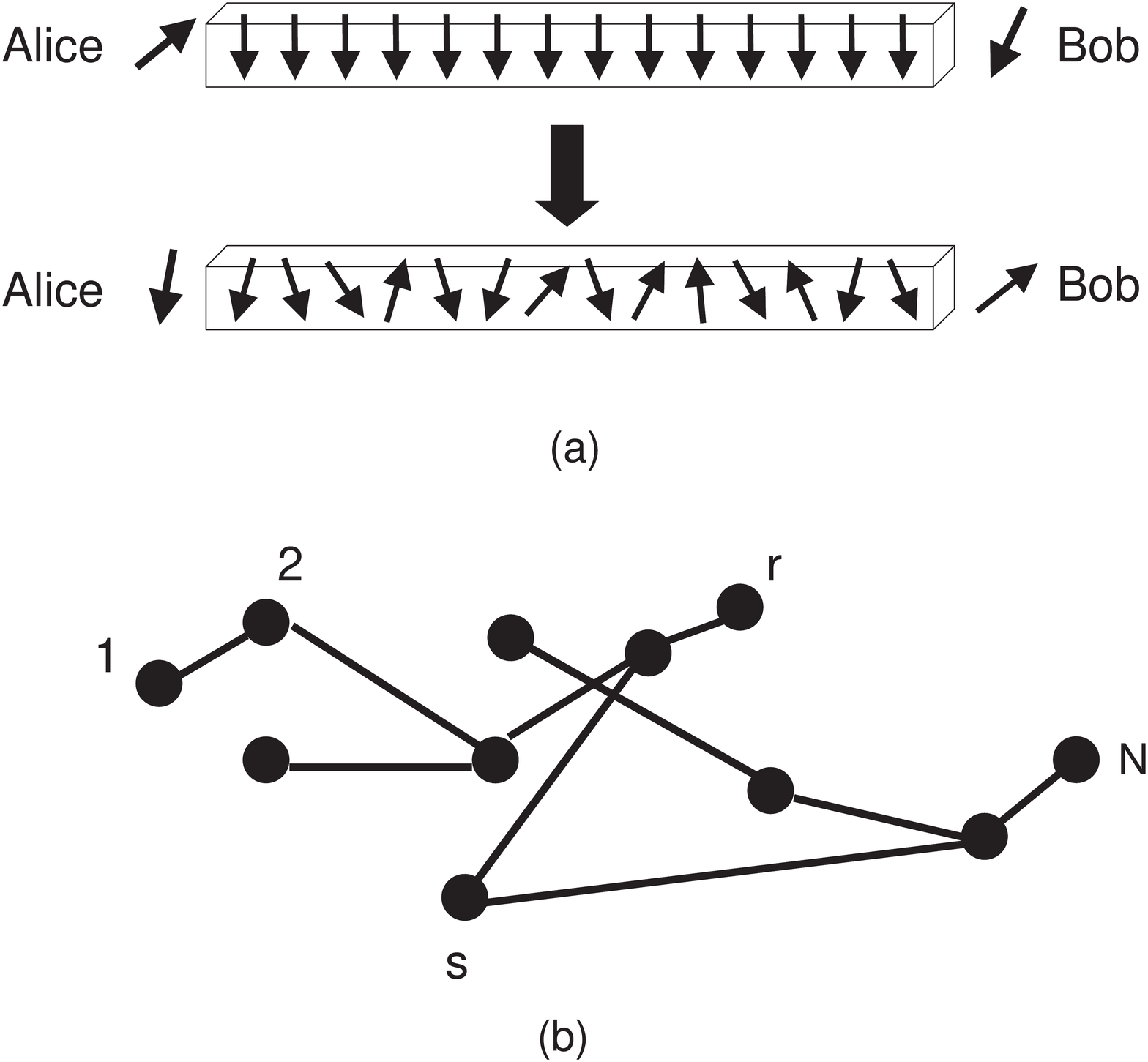}
 \caption{The part (a) of the figure shows our quantum communication protocol. Initially the spin chain is in
 its ground state in an external magnetic field. Alice and Bob are at opposite ends of the chain.
 Alice places the quantum state she wants to communicate on the spin nearest to her. After a while, Bob receives this
 state with some fidelity on the spin nearest to him. Part (b) shows an arbitrary graph of spins through which quantum
 communications may be accomplished using our protocol. The communication takes place from the sender spin $s$ to the receiver
 spin $r$.}
\label{QT1}
\end{center}
\end{figure}

  Transmitting a quantum state (known or unknown) from one place to another is often
an important task \cite{bennett00}. It is required, for example,
to link several small quantum processors for large-scale quantum
computing. Thus it is very important to have physical systems
which can serve as {\em channels} for quantum communication. We
can either directly transmit a state through the channel, or we
can first use the channel to share entanglement with a separated
party and then use this entanglement for teleportation
\cite{bennett93}. The ideal channel for long distance quantum
communications is an optical fiber. This requires interfacing a
quantum computer (such as arrays of spins or ions) with optics.
For short distance communications (say between adjacent quantum
processors), alternatives to interfacing different kinds of
physical systems are highly desirable and have been proposed, for
example, for ion traps \cite{wineland02}. In this letter, I
propose a scheme to use a spin chain (a 1D magnet -- real or
simulated) as a channel for short distance quantum communication.
The communication is achieved by placing a spin encoding the state
at one end of the chain and waiting for a specific amount of time
to let this state propagate to the other end (as shown in
Fig.\ref{QT1}(a)). This helps to avoid interfacing because both
quantum computers and quantum channels can then be made by the
same physical systems. Moreover, neither does the spin chain
channel require the ability to switch interactions  "on" and "off"
(often a problem in quantum computer implementations
\cite{zhou,benjamin}), nor does it require {\em any} modulation by
external fields (essential for quantum computation). This
simplicity in comparison to a quantum computer makes it an ideal
connector between quantum computers and realizable well before a
quantum computer.

       I will first present the scheme in a general
setting for arbitrary graphs of spins with ferromagnetic
Heisenberg interactions and later proceed to the realizable case
of an open ended chain. Consider the general graph shown in
Fig.\ref{QT1}(b), where the vertices are spins and the edges
connect spins which interact. Say there are $N$ spins in the graph
and these are numbered $1,2,...,N$. The Hamiltonian is given by
\begin{equation}
{\bf H}_{\bf G}=-\sum_{<i,j>}
J_{ij}~\vec{\sigma}^i.\vec{\sigma}^{j}-\sum_{i=1}^N B_i\sigma_z^i.
\label{ham}
\end{equation}
$\vec{\sigma}^i=(\sigma_x^i, \sigma_y^i, \sigma_z^i)$ in which
$\sigma_{x/y/z}^i$ are the Pauli matrices for the $i$th spin,
$B_i>0$ are {\em static} magnetic fields and $J_{ij}>0$ are
coupling strengths, and $<i,j>$ represents pairs of spins. ${\bf
H}_{\bf G}$ describes an arbitrary ferromagnet with isotropic
Heisenberg interactions. We now assume that the state sender Alice
is located closest to the $s$th ({\em sender}) spin and the state
receiver Bob is located closest to the $r$th ({\em receiver}) spin
(these spins are shown in Fig.\ref{QT1}(b)). All the other spins
will be called {\em channel spins}. It is also assumed that the
sender and receiver spins are {\em detachable} from the chain. In
order to transfer an unknown state to Bob, Alice replaces the
existing sender spin with a spin encoding the state to be
transferred. After waiting for a specific amount of time, the
unknown state placed by Alice travels to the receiver spin with
some fidelity. Bob then picks up the receiver spin to obtain a
state close to the the state Alice wanted to transfer. As we never
require individual access or individual modulation of the channel
spins in our protocol, they can be constituents of rigid 1D
magnets.

We assume that initially the system is initially cooled to its
ground state $|{\bf 0}\rangle=|000...0\rangle$ where $|0\rangle$
denotes the spin down state (i.e., spin aligned along $-z$
direction) of a spin. This is shown for a 1D chain in the upper
part of Fig.\ref{QT1}(a). I will set the ground state energy
$E_0=0$ ({\em i.e.,} redefine ${\bf H}_{\bf G}$ as $E_0+{\bf
H}_{\bf G}$) for the rest of this paper. We also introduce the
class of states $|{\bf j}\rangle=|00...010....0\rangle$ (where
${\bf j}={\bf 1},{\bf 2},..{\bf s},..{\bf r},..,{\bf N}$) in which
the spin at the $j$th site has been flipped to the $|1\rangle$
state. To start the protocol, Alice places a spin in the unknown
state$|\psi_{in}\rangle=
\cos{(\theta/2)}|0\rangle+e^{i\phi}\sin{(\theta/2)}|1\rangle$ at
the $s$th site in the spin chain. We can describe the state of the
whole chain at this instant (time $t=0$) as
\begin{equation}
|\Psi(0)\rangle=\cos{\frac{\theta}{2}}|{\bf
0}\rangle+e^{i\phi}\sin{\frac{\theta}{2}}|{\bf s}\rangle.
\end{equation}
Bob wants to retrieve this state, or a state as close to it as
possible, from the $r$th site of the graph. Then he has to wait
for a specific time till the initial state $|\Psi(0)\rangle$
evolves to a final state which is as close as possible to
$\cos{\frac{\theta}{2}}|{\bf
0}\rangle+e^{i\phi}\sin{\frac{\theta}{2}}|{\bf
   r}\rangle$. As $[{\bf H}_{\bf G},\sum_{i=1}^N \sigma_z^i]=0$, the state $|{\bf
s}\rangle$ only evolves to states $|{\bf j}\rangle$ and the
evolution of the spin-graph (with $\hbar=1$) is
\begin{equation}
|\Psi(t)\rangle=\cos{\frac{\theta}{2}}|{\bf
0}\rangle+e^{i\phi}\sin{\frac{\theta}{2}}\sum_{{\bf j}={\bf
1}}^{\bf N} \langle {\bf j}|e^{-i{\bf H}_{\bf G}t}|{\bf
s}\rangle|{\bf j}\rangle.
\end{equation}
The state of the $r$th spin will, in general, be a mixed state,
and can be obtained by tracing off the states of all other spins
from $|\Psi(t)\rangle$. This evolves with time as
\begin{equation}
\rho_{out}(t)=P(t)|\psi_{out}(t)\rangle\langle\psi_{out}(t)|+(1-P(t))|0\rangle\langle
0|,
\label{out}
\end{equation}
with
\begin{equation}
|\psi_{out}(t)\rangle =
\frac{1}{\sqrt{P(t)}}(\cos{\frac{\theta}{2}}|
0\rangle+e^{i\phi}\sin{\frac{\theta}{2}}
f^N_{s,r}(t)|1\rangle),\label{out3}
\end{equation}
where
$P(t)=\cos^2{\frac{\theta}{2}}+\sin^2{\frac{\theta}{2}}|f^N_{r,s}(t)|^2$
and $f^N_{r,s}(t)=\langle {\bf r}|\exp{\{-i{\bf H}_{\bf
G}t\}}|{\bf s}\rangle$. Note that $f^N_{r,s}(t)$ is just the
transition amplitude of an excitation (the $|1\rangle$ state) from
the $s$th to the $r$th site of a graph of $N$ spins.

     Now suppose it is decided that Bob will pick up the
$r$th spin (and hence complete the communication protocol) at a
predetermined time $t=t_0$. The fidelity of quantum communication
through the channel averaged over all pure input states
$|\psi_{in}\rangle$ in the Bloch-sphere ($(1/4\pi)\int
\langle\psi_{in}| \rho_{out}(t_0)|\psi_{in}\rangle d\Omega$) is
then
\begin{equation}
F=\frac{|f^N_{r,s}(t_0)|\cos{\gamma}}{3}+\frac{|f^N_{r,s}(t_0)|^2}{6}+\frac{1}{2},
\label{fid}
\end{equation}
where $\gamma=\arg\{f^N_{r,s}(t_0)\}$. To maximize the above
average fidelity, we must choose the magnetic fields $B_i$ such
that $\gamma$  is a multiple of $2\pi$. Assuming this special
choice of magnetic field value (which can always be made for any
given $t_0$) to be a part of our protocol, we can simply replace
$f^N_{r,s}(t_0)$ by $|f^N_{N,1}(t_0)|$ in Eq.(\ref{out3}). The
spin chain then acts as an {\em amplitude damping quantum channel}
\cite{preskill-notes,nielsen-book}. It converts the input state
$\rho_{in}=|\psi_{in}\rangle\langle\psi_{in}|$ to
$\rho_{out}=M_0\rho_{in}M_0^\dagger+M_1\rho_{in}M_1^\dagger$ with
the operators $M_0$ and $M_1$ (Kraus operators
\cite{preskill-notes}) given by
\begin{equation}
M_0=\left[\begin{array}{ll} 1 & ~~~0 \\
0 & |f^N_{r,s}(t_0)|
\end{array}\right], M_1=\left[\begin{array}{ll} 0 & \sqrt{1-|f^N_{r,s}(t_0))|^2} \\
0 & ~~~~~~~~~~0
\end{array}\right].
\end{equation}

   Now consider the transmission of the state of one member of a pair of particles in the entangled state
$|\psi^{+}\rangle=\frac{1}{\sqrt 2}(|0 1\rangle+|1 0\rangle)$
through the channel. This is the usual procedure for sharing
entanglement between separated parties which {\em precedes}
teleportation \cite{bennett00}. The output state $\rho_{out}(t_0)=
\sum_{i=0,1}I\otimes M_i|\psi^{+}\rangle\langle\psi^{+}| I\otimes
M_i^\dagger$ is
\begin{eqnarray}
\rho_{out}(t_0)&=&\frac{1}{2}\{(1-|f^N_{r,s}(t_0)|^2)|00\rangle\langle00|\nonumber\\
&+&(|10\rangle+|f^N_{r,s}(t_0)||01\rangle)(\langle
10|+|f^N_{r,s}(t_0)|\langle 01|)\}\nonumber
\end{eqnarray}
The entanglement ${\cal E}$ of the above state, as measured by the
concurrence \cite{wootters} is given by
\begin{equation}
{\cal E}=|f^N_{r,s}(t_0)|. \label{ent}
\end{equation}
Thus, for any non-zero $f^N_{r,s}(t_0)$ (however small), some
entanglement can be shared through the channel. This entanglement,
being that of a $2\times 2$ system, can also be {\em distilled}
\cite{horodecki} into pure singlets and used for teleportation.
Later we will estimate $f^N_{r,s}(t_0)$ for very long open chains
and show that entanglement can be distributed to arbitrary
distances.

 Eqs.(\ref{fid}) and (\ref{ent}) are exceptionally
simple formulas for the fidelity of quantum communications and the
entanglement shared through our spin-graph channel in terms of
{\em single} transition amplitude $f^N_{r,s}(t_0)$. We note here
that such simple formulas, with slight modifications, will hold
for spin-graphs with much wider class of interactions, as long as
the state $|{\bf 0}\rangle$ does not evolve \cite{nielsen1}.

    We will now consider a linear open ended spin chain (Fig.\ref{QT1}(a)),
which is the most natural geometry for a channel. To use an
analytically solvable Hamiltonian ${\bf H}_{\bf L}$ we assume
$J_{ij}=(J/2)\delta_{i+1,j}$ (nearest neighbor interactions of
equal strength) and $B_i=B$ (uniform magnetic field) for all $i$
and $j$ in Eq.(\ref{ham}) for ${\bf H}_{\bf G}$. The eigenstates
of ${\bf H}_{\bf L}$, relevant to our problem are
\begin{equation}
|\tilde{m}\rangle_{\bf L}=a_m\sum_{j=1}^N
\cos\{{\frac{\pi}{2N}(m-1)(2j-1)}\}|{\bf j}\rangle,
\end{equation}
where $m=1,2,...,N$, $a_1=1/\sqrt{N}$ and $a_{m\neq 1}=\sqrt{2/N}$
with energy (on setting $E_0=0$) given by
$E_m=2B+2J(1-\cos\{{\frac{\pi}{N}(m-1)}\})$. In this case,
$f^N_{r,s}(t_0)$ is given by
\begin{equation}
f^N_{r,s}(t_0) = \sum_{m=1}^N \langle {\bf N}|\tilde{m}\rangle
\langle \tilde{m}|{\bf 1}\rangle e^{-iE_mt_0} = IDCT_s(v_m)
\label{dct}
\end{equation}
where, $v_m=a_m\cos{\{\frac{\pi}{2N}(m-1)(2r-1)\}}e^{-iE_mt_0}$
and $IDCT_s(v_m)=\sum_{m=1}^N a_m v_m
\cos{\{\frac{\pi}{2N}(m-1)(2s-1)\}}$ is the $s$th element of the
inverse discrete cosine transform of the vector $\{v_m\}$.

\begin{figure}
\begin{center}
\includegraphics[width=3in, clip]{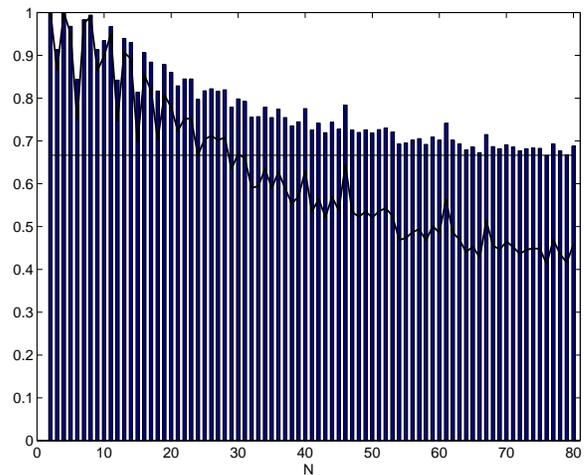}
 \caption{The bar plot shows the maximum
fidelity $F$ of quantum communication and the curve shows the
maximum sharable entanglement ${\cal E}$ achieved in a time
interval $[0,4000/J]$ as a function of the chain length $N$ from
$2$ to $80$. The time $t_0$ at which this maxima is achieved
varies with $N$. The straight line at $F=2/3$ shows the highest
fidelity for classical transmission of a quantum state.}
\label{QT3}
\end{center}
\end{figure}

\begin{figure}
\begin{center}
\includegraphics[width=3in, clip]{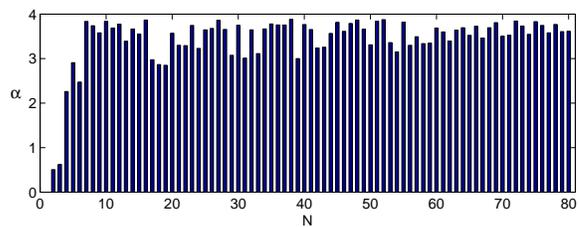}
 \caption{The upper part plots $\alpha=\log_{10}{2Jt_0}$  where $t_0$ is a time  in the interval
 $[0,T_{max}=4000/J]$ when the maximum
fidelities $F$ plotted in fig\ref{QT3} are achieved.}
\label{QT4}
\end{center}
\end{figure}

 We now want to study the performance of our protocol for various
chain lengths $N$ with $s=1$ and $r=N$ (Alice and Bob at opposite
ends of the chain as shown in Fig.\ref{QT1}(a)). Bob has to wait
for different lengths of time $t_0$ for different chain lengths
$N$, in order to obtain a high fidelity of quantum state transfer.
Using Eqs.(\ref{fid}),(\ref{ent}) and (\ref{dct}), I have
numerically evaluated the maximum of $|f^N_{N,1}(t_0)|$ (which
corresponds to the maxima of both fidelity and entanglement) for
various chain lengths from $N=2$ to $N=80$ when Bob is allowed to
choose $t_0$ within a finite (but long) time interval of length
$T_{max}=4000/J$. This evaluation is fast because Eq.(\ref{dct})
allows us to use numerical packages for the discrete cosine
transform. Taking a finite $T_{max}$ is physically reasonable, as
Bob cannot afford to wait indefinitely. It is to be understood
that within $[0,T_{max}]$, the time $t_0$ at which optimal quantum
communication occurs varies with $N$. The maximum fidelities  as a
function of $N$  and the maximum amounts of entanglement sharable
(both rounded to $3$ decimal places) are shown in Fig.\ref{QT3}.
The corresponding times $t_0$ are shown as a logarithmic plot in
Fig.\ref{QT4}. Note that the search for $t_0$ is numerical, so its
value may be any one of several instances of time when the maxima
of $|f^N_{N,1}(t_0)|$ (to $3$ decimal places) in the range
$[0,T_{max}]$ is attained ($t_0$ is not necessarily the least time
at which the plotted fidelities are achieved).

Fig.\ref{QT3}, shows various interesting features of our protocol.
The plot also shows that in addition to $N=2$, which is perfect (a
well known fact for the Heisenberg interaction \cite{divincenzo}),
$N=4$ gives perfect ($F=1.000$) quantum state transfer to $3$
decimal places and $N=8$ gives near perfect ($F=0.994$). The
fidelity also exceeds $0.9$ for $N=7,10,11,13$ and $14$. Till
$N=21$ we observe that the fidelities are lower when $N$ is
divisible by $3$ in comparison to the fidelities for $N+1$ and
$N+2$. The plot also shows that a chain of $N$ as high as $80$
exceeds the highest fidelity for classical transmission of the
state i.e., $2/3$ \cite{horodecki1} in the time interval probed by
us. Of course the above results hold {\em only} when one is trying
to directly transmit the quantum state over a distance. If one
first shares entanglement through the channel, then the amount of
entanglement $\cal{E}$ is about $0.45$ for an $80$ spin chain.
This entanglement can be distilled to pure singlets and used for
perfect teleportation.

  We will now like to estimate the entanglement sharable through chains so large that it is difficult to
identify an optimal $t_0$ by numerical search. Hence we will
choose $t_0$ according to a fixed (in general, non-optimal)
prescription. To motivate this choice, we expand $e^{-iE_mt_0}$ in
Eq.(\ref{dct}) as a Bessel function series to obtain
\begin{equation}
{\cal E}=|\sum_{k=-\infty}^\infty (-1)^k
(J_{N+Nk}(\beta_0)+iJ^{'}_{N+Nk}(\beta_0))|, \label{line}
\end{equation}
where $\beta_0=2Jt_0$. Using, $J_N(N+\xi
N^{1/3})\approx(2/N)^{1/3}Ai(-2^{1/3}\xi)$ for large $N$
\cite{childs}, (where $Ai(.)$ is the Airy function) we can prove
that we get a maxima of $J_N(\beta_0)$ at $t_0=(N+0.8089
N^{1/3})/2J$ and at this time
\begin{equation}
{\cal E}\approx 2|J_N(\beta_0)|\approx 1.3499 N^{-1/3},
\end{equation}
which ranges from $0.135$ for $N=1000$ to $1.35\times10^{-4}$ for
$N=10^{12}$ (just $3$ orders decrease in ${\cal E}$ for an
increase in length $N$ by $9$ orders -- a very efficient way to
distribute entanglement). Thus for {\em any} finite $N$, however
large, the chain allows us to distribute entanglement of the order
$N^{-1/3}$ in a time $t_0$ linear in $N$.

    As an alternate system, we now consider a ring
of $2N$ spins with Hamiltonian ${\bf H}_{\bf R}$ obtained by using
$J_{ij}=(J/2)\delta_{i\oplus 1,j},B_i=B$ in
Eq.(\ref{ham})($\oplus$ is summation modulo $2N$). Alice and Bob
access the spins at diametrically opposite sites ($s=1$, $r=N+1$).
In this case, the $2N$ eigenstates in the one excitation sector
are $|\tilde{m}\rangle_{\bf R}=(1/\sqrt{2N})\sum_{j=1}^{2N} e^{i
\frac{\pi}{N}(m-1) j}|{\bf j}\rangle$ and
\begin{equation}
{\cal E}=|IDFT_{r-s}(u_m)|=|\sum_{k=-\infty}^\infty (-1)^k
J_{N+Nk}(\beta_0)|, \label{ring}
\end{equation}
where $u_m=\exp(-iE_m t_0)$ and
$IDFT_{r-s}(u_m)=(1/2N)\sum_{m=1}^{2N} u_m
\exp\{i(2\pi/2N)(r-s)(m-1)\}$ is the $(r-s)$th component of the
inverse discrete fourier transform of the vector $\{u_m\}$.
Eqs.(\ref{line}) and (\ref{ring}), and the nature of Bessel
functions, imply that the maxima of ${\cal E}$ coincide for the
line and the ring. This means that by using a ring you can
communicate as efficiently over a distance $r-s=N$ as you can with
a open ended line over a distance $r-s=N-1$. An immediate
implication is that a $4$ spin ring allows perfect communication
between diametrically opposite sites (because a $2$ spin line does
\cite{divincenzo}). This simple result in quantum information was
not known till now. The coincidence of the maxima also means that
the maxima of ${\cal E}$ for the line can also be computed by
inverse fourier transforming $\{u_m\}$.

   We now mention potential systems for realization.
Josephson junction arrays, excitons in quantum dots and real 1D
magnets, which motivated the recent study of quantum computation
in Heisenberg chains \cite{zhou,benjamin} will be good candidates.
Interactions in such systems are difficult to tune. Our scheme can
be implemented in such systems without the elaborate control
required for quantum computation. 1D arrays of spins in solids
\cite{divincenzo,kane} are also candidates. There are ring
molecules described exceptionally well by ${\bf H}_{\bf R}$, which
also allow local probes for individual spins \cite{loss} (these
are antiferromagnetic, but a large $B$ could make $|{\bf
0}\rangle$ the ground, and $|\tilde{m}\rangle_{\bf R}$ the first
excited states). Benzene molecules (with NMR probes possible) with
$J_{ij}=\frac{\delta_{i\oplus 1,j}}{4}+\frac{\delta_{i\oplus
2,j}}{12\sqrt{3}}+\frac{\delta_{i\oplus 3,j}}{32}$, still have
$|\tilde{m}\rangle_{\bf R}$ as eigenstates \cite{bowden}. $F$ can
thus be calculated by an IDFT to be $0.793$ for $r-s=3$ at
$t_0=130$. Principles of the scheme should also be testable in
simulated open ended Heisenberg chains in a 1D optical lattice
\cite{duan}.

   In this letter, I have presented a protocol for quantum
communication through an unmeasured and unmodulated spin chain. It
allows quantum communication between adjacent quantum computers
without interfacing different physical systems. It is well known
that there exists an alternate trivial method of transferring
quantum states perfectly over a distance by a series of swaps. But
that requires an elaborate sequence of time dependent fields. The
highly non-trivial finding of this letter is that even without
doing anything, simply by placement, quantum states can be
transmitted with high fidelity over a significant distance and
entanglement of the order $N^{-1/3}$ can be shared across a chain
of length $N$. We also found that a $4$ spin ring allows perfect
quantum communication between diametrically opposite sites. This
letter can be regarded as a study of a fundamental condensed
matter system (a finite ferromagnet and its excitations) from the
viewpoint of quantum communications. There remains an enormous
scope for future extensions to spin graphs of varied geometry and
interactions and to other well known condensed matter systems.

   This work is supported by the NSF under Grant Number
   EIA-00860368. I thank G. J. Bowden,  J. Eisert, A. Ekert, J. Harrington, V. Korepin,  R.
   Raussendorf,
   A. Thapliyal, F. Verstraete and particularly Michael Nielsen and John Preskill, for remarks and suggestions.




\end{document}